# Field induced changes in cycloidal spin ordering and coincidence between magnetic and electric anomalies in BiFeO$_3$ multiferroic

B. Andrzejewski[1], A. Molak[2], B. Hilczer[1], A. Budziak[3], R. Bujakiewicz-Korońska[4]

[1] Institute of Molecular Physics, PAN, Smoluchowskiego 17, PL-60179 Poznań, Poland

[2] University of Silesia in Katowice, Institute of Physics, Bankowa 14, PL-40007 Katowice, Poland

[3] The H. Niewodniczański Institute of Nuclear Physics, PAN, Radzikowskiego 152, PL-31342 Cracow, Poland

[4] Pedagogical University in Cracow, Institute of Physics, Podchorążych 2, PL-30084 Cracow, Poland

Corresponding author: B. Andrzejewski, e-mail: Bartlomiej.Andrzejewski@ifmpan.poznan.pl

**Abstract**

The ZFC and FC magnetization dependence on temperature was measured for BiFeO$_3$ ceramics at the applied magnetic field up to $\mu_0 H = 10$ T in 2 K ÷ 1000 K range. The antiferromagnetic order was detected from the hysteresis loops below the Neel temperature $T_N = 646$ K. In the low magnetic field range there is an anomaly in $M(H)$, probably due to the field-induced transition from circular cycloid to the anharmonic cycloid. At high field limit we observe the field-induced transition to the homogeneous spin order. From the $M(H)$ dependence we deduce that above the field $H_a$ the spin cycloid becomes anharmonic which causes nonlinear magnetization, and above the field $H_c$ the cycloid vanishes and the system again exhibits linear magnetization $M(H)$. The anomalies in the electric properties, $\varepsilon'(T)$, $\tan\delta(T)$, and $\sigma(T)$, which are manifested within the 640 K ÷ 680 K range, coincide to the anomaly in the magnetization $M(T)$ dependence, which occurs in the vicinity of $T_N = 646$ K. We propose to ascribe this coincidence to the critical behaviour of the chemical potential $\mu_S$, related to the magnetic phase transition.

# 1. INTRODUCTION

Bismuth ferrite $BiFeO_3$ (BFO) belongs to multiferroic compounds. It exhibits a coexistence of antiferromagnetic and ferroelectric ordering that attracts attention due to possible applications in sensors and multi-state memory devices [1-5]. The intrinsic properties of the BFO have been determined for single crystal samples [4, 6]. It shows a rhomboedrally distorted perovskite cell with space group *R3c* at room temperature.

The G-type antiferromagnetic (AFM) order below temperature $T_N$ = 643 K with superimposed long-range incommensurate cycloidal spiral structure was reported [7]. However, there are controversial reports also, e.g. a lack of cycloidal structure influence on AFM susceptibility was noticed in the literature. Despite the weak ferromagnetism was detected in bismuth ferrite [8], actually it may be also ascribed to magnetic properties of the precipitated impurity phase since this type of ordering vanished in the pure BFO [4, 9].

The magnetic order in BFO is a subject to several transitions manifesting in magnetic properties near the temperatures 140 K, 200 K and 643 K. The last and most distinct one is the transition between antiferromagnetic and paramagnetic order (AFM-PM transition). The transition at 140 K is interpreted in terms of spin-glass [10] and the anomaly about 200 K is ascribed to magnon softening [10].

The transitions in magnetic ordering can be also induced by means of an external magnetic field. A homogeneous AFM spin state with no superimposed cycloidal spin ordering in bismuth ferrite can be obtained under the influence of a very strong magnetic field exceeding 20 T and at room temperature [9].

The ferroelectric features, with the Curie temperature $T_C$ = 1143 K, were confirmed by the *P-E* hysteresis loop measurement with a remnant polarisation $P_r \approx 35$ µC cm$^{-2}$ at room temperature [4]. The measured ferroelectric remnant polarisation $P_r \sim 35$ µC cm$^{-2}$ is close to the value estimated from first-principles calculations [11, 12]. Moreover, this quantity value and the electric polarisation *P(E)* hysteresis loop depend markedly on the oxygen vacancies content [13].

The coexistence and interaction of the antiferromagnetic and ferroelectric phases is confirmed both in crystals and ceramics BFO [4, 14]. The influences of the applied electric field *E* on the magnetic magnetisation *M* and the applied magnetic field *H* on the electric polarization *P* have been reported. However, the significant difference between the phase transition temperatures, i.e. $T_N \approx 643$ K for the AFM-PM transition and the $T_C \approx 1143$ K for the FE-PE transition, suggest that direct interaction between the magnetic and electric subsystem might be week.

The stoichiometric BiFeO$_3$ is a highly resistive material and exhibits the resistivity of the GΩm order at room temperature [4]. The calculated energy gap in the electronic structure is equal to 2.8 eV [15]. The optical energy gap was estimated as 2.7 eV [16]. However, the electric conduction activation energy $E_a$ of the BFO ceramics was reported as varying e.g. from 0.2, 0.3, 0.58, 0.9, to 1.1 eV [17-19], dependably on temperature range. Hence, the electric conduction depends on the crystal lattice defects, both the oxygen vacancies and doping, e.g. with Nb [17], La and Mn [20, 21] which influence the electrical conductivity value by several orders in magnitude. It is worthwhile to note that the transition from states ~0.3 eV below the band gap were related to the oxygen vacancies, basing on the refractivity measurements [16].

The aims of the work were: (1) to determine the magnetisation $M(H)$ and cycloidal spin ordering dependence on strong magnetic fields; (2) to determine whether the magnetic AFM-PM phase transition can be detected by means of the electrical measurements.

## 2. EXPERIMENTAL METHODS

### 2.1. Sample preparation

The BiFeO$_3$ ceramics were obtained by the hot pressing method. The Bi$_2$O$_3$ and Fe$_2$O$_3$ starting materials were mechano-activated when milled for 24 h in a ball mill filled with ethanol, then calcined at 923 K. The product was crushed, grinded, and then sintered at 1123 K. The proper composition was confirmed with the EDS test [22].

### 2.2 Magnetic test

The magnetometric measurements were performed by means of Quantum Design PPMS System (Physical Property Measurement System). Two magnetometer probes fitted to PPMS were used. The first one was extraction magnetometer (ACMS) probe allowing both DC magnetic moment and AC susceptibility measurements and the second was VSM (Vibrating Sample Magnetometer) probe equipped additionally with high temperature set. The BiFeO$_3$ sample, for magnetic test, was cut to proper dimensions (about 3x3x5 mm$^3$). It was mounted in a standard PTFE straw when the ACMS probe was used and it was mounted in a bronze sample support for the measurements performed by means of VSM probe. For high temperature measurements the thin slab (about 3x3x1mm$^3$) was cut from BiFeO$_3$ sample to avoid of any temperature gradients.

### 2.3 X-ray diffraction test

X-ray diffraction (XRD) studies have been carried out using an X'Pert PRO (PANalytical) diffractometer with the Cu K$_\alpha$ radiation and a graphite monochromator. Temperature of a sample has been stabilized (accuracy ±1 K) with the high-temperature oven-chamber HTK 1200N supplied by an Anton Paar Co. The BiFeO$_3$ powder sample was placed in a ceramic container (the diameter of 16 mm and the depth of ca. ~0.8 mm). No specimen was added to the sample to stick it to the container. Before the measurement, the chamber was pumped-out to pressure about 1 mbar. The measurements have been performed during heating. After each heating stage the sample has been held about 10 min to obtain equilibrium. A profile-fitting program FULLPROF [23] based on the Rietveld method was used to analyze and fit the XRD patterns.

### 2.4 Electrical test

A sample in the form of parallelepiped (about 2x3x1 mm$^3$) was cut from the sintered ceramic pellets. The surfaces were polished with a diamond paste, grade 2000. The samples for electric measurements were polished and Ag paste (Leitsilber 200, Hans Wolbring GmbH) electrodes were painted and fired. The $C_p(T,f)$ and $G(T,f)$ was measured with a use of a HP4263B LCR meter. The effective quantities $\varepsilon'(T)$, $\tan\delta(T)$, and $\sigma(T)$ were evaluated assuming a standard dielectric capacitor model. Temperature was varied within the range 290 ÷ 700 K, at a rate ±3 K/min.

### 3. RESULTS AND DISCUSSION
### 3.1 Magnetic properties

The magnetization dependence on temperature $M(T)$ for BiFeO$_3$ sample studied for the applied magnetic field $\mu_0H$ = 1 T is shown in Figure 1. The magnetization curves measured after zero field cooling (ZFC) and after field cooling (FC) in the field $\mu_0H$ = 1 T are exactly the same which proves that there irreversible processes in the sample are negligible. This observation is in contradiction to the results presented in a work of Singh et al. [10] who reported substantial irreversibility between ZFC and FC magnetization for the same applied magnetic field value as in our case. They explained this irreversible effect in terms of spin-glass behavior.

The magnetization $M(T)$ rapidly decreases on heating in the low temperature range (2 K ÷ ~50 K), it exhibits a plateau within a range 50 K ÷ 150 K, and it gradually increases at higher temperatures (150 K ÷ ~640 K). This type of $M(T)$ dependence is consistent with the previous reports [10, 24, 25]. The maximum in $M(T)$ observed in the low temperature range cannot be explained in terms of changes in population or domain reorientations because no sudden changes in domain structure occur at this temperature range [24]. Hence, we deduce that the marked decrease in the $M(T)$ value within the low temperature range results from small amounts of ferromagnetic contaminations.

The AC susceptibility measurement, shown in Fig. 2, confirms the magnetization measurement result, i.e. the real part of susceptibility $\chi'(T)$ is identical to magnetization dependence on temperature $M(T)$. The values of the imaginary part $\chi''(T)$ are too low to deduce any conclusions. However, this observation once again proves that the irreversible magnetization mechanisms are absent in our sample.

Figure 3 presents several magnetization loops $M(H)$ obtained for the $BiFeO_3$ ceramics, recorded for chosen temperatures from 4 K to 300 K range. They exhibit no magnetic irreversibility in agreement with ZFC-FC magnetization and AC measurements, and almost the linear dependence of magnetization $M(H)$ on the applied magnetic field (-2T< $\mu_0H$< +2T). Such behavior of magnetization is consistent with the antiferromagnetic ordering in bismuth ferrite [26-29].

Significant results of magnetometric measurements appear in the high temperature range, close to the Néel temperature, at the transition to paramagnetic state, which are shown in Fig. 4. The magnetic moment $M(T)$ increases with temperature and exhibits a marked peak ascribed to the phase transition temperature. The phase transition occurs at $T_N = 646$ K (when measured in the middle of the magnetization maximum) and its width is 8 K. The behavior of the magnetization near the Néel temperature $T_N$ can be explained in terms of changes in ratio of the population of the in-plane magnetic domains to the out-of-plane ones. It has been reported that this ratio decreases rapidly from almost constant value equal to 1.6 in a wide temperature range to about 1.0 near the AFM phase transition and can be related to a strong increase in the magnetization [24]. In our case, the ratio value by which the magnetization drops at the Néel temperature is equal to 1.4 that seems to be comparable to the literature report change in ratio of the populations of domains which is 1.6.

Two magnetization curves vs. magnetic field *M(H)* obtained at temperature 304 K and 577 K are shown in Fig. 5. For the temperature equal to 304 K, the *M(H)* exhibits the nonlinearity in a very narrow field range when the $\mu_0 H$ value *is* close to zero. Then, the *M(H)* increases linearly when the magnetic field increases and finally the *M(H)* deviates from linear dependence above a field $H_a$ (noted in Fig. 5). Another dependence occurs in the *M(H)* curve recorded for temperature equal to 577 K. Neglecting the low magnetic fields close to zero, the magnetization increases linearly. It exhibits non-linear changes in the intermediate field range, i.e. between the fields $H_a$ (i.e. anharmonic) and $H_c$ (i.e. critical), and it increases once again linearly at high fields above $H_c$.

Kadomtseva et al. [30] proposed to interpret the origin of the field $H_c$ followed by the linear *M(H)* magnetization in high magnetic field in terms of the field induced phase transitions. Namely, at low magnetic fields the spin cycloid is nearly circular (Fig. 6a), at intermediate fields the cycloid becomes anharmonic due to interaction of spins system with the magnetic field (Fig. 6b), and at high fields $H > H_c$ there is a transition to the homogeneous spin order (Fig. 6c). Therefore, in the field strong enough, $H > H_c$, the system of spins in the cycloid is aligned with magnetic field direction and exhibits a linear magnetization. Such a transition has been already observed in magnetization, dielectric and ESR investigations [25, 30-33]. We assume that, for the room temperature, the $H_c$ appears in too high magnetic fields (estimated as about 15 T) to be recorded by PPMS System equipped with 9 T magnet, that explains the absence of this transition in *M(H)* curve measured at 304 K.

The temperature dependence of the magnetization *M(H)* is related to the cycloidal order of Fe spins and has been already modeled by Kadomtseva et al. [30] and by Popov et al. [33]. In the case of the field applied in [001]$_c$ direction, the component of the magnetization $M_{[001]}(H)$ along the field is as follows [30]:

$$M_{[001]}(H) = M_{[001]}^S \langle sin\theta \rangle_\lambda + \frac{2}{3}\chi_\perp H + \frac{1}{3}\chi_\perp H \langle sin^2\theta \rangle_\lambda \qquad (1)$$

where $\theta$ is the angle between antiferromagnetic vector and the vector of the spontaneous magnetization $P_s$ (hexagonal c-axis or cubic [111]$_c$ direction).

For the low magnetic fields we have $<\sin\theta>_\lambda \approx 0$, and $<\sin^2\theta>_\lambda \approx \frac{1}{2}$, hence $M_{[001]}(H)$ takes the form $M_{[001]}(H) = 5/6\chi_\perp H$ dependence. Above the critical field $H_c$, the spin cycloid disappears and homogeneous magnetic ordering is established, which corresponds to $<\sin\theta>_\lambda \approx 1$, $<\sin^2\theta>_\lambda \approx 1$, and the eq. (1) transforms to $M = M^S_{[001]}+\chi_\perp H$ [30].

The mean magnetization of the polycrystalline sample $\langle M(H)\rangle$ along magnetic field, i.e. in our case measurement, is obtained from average of eq. (1) taken over all orientations of the applied magnetic field $H$ with respect to $P_s$ vector of each individual crystalline grain, therefore: $\langle M(H)\rangle=2/\pi M_{[001]}(H)$.

It has been observed and reported that the model equation (eq.1) roughly fitted experimental data for single crystals of $BiFeO_3$ [30, 34]. The identical disagreement is visible in Fig. 5, which shows magnetization data for the ceramic, polycrystalline $BiFeO_3$ sample. The following values of parameters: $\mu_0H_c = 8.25$ T, $\chi_\perp = 1.28 \ 10^{-5}$ emu/g·Oe, $M_s = 0,199$ emu/g has been obtained with the best numerical fitting procedure. These values can be compared to the values of parameters determined for the single crystal: $\mu_0H_c \approx 20$ T, $\chi_\perp \approx 0,6 \ 10^{-5}$ emu/g·Oe, and $M_s \approx 0,25$ emu/g [35]. However, our measurements were performed at much higher temperature $T = 577$ K than in the case of the single crystals, which were investigated at $T = 10$ K [35]. At high temperature as in our case one can expect that the critical field $H_c$ and the spontaneous magnetization $M_s$ decreases but the susceptibility $\chi_\perp$ increases because of the lower "stiffness" of the spin cycloid. The increased value of $\chi_\perp$ at the high temperature corresponds also well to the rise of magnetization observed close to the Neel temperature at the AFM-PM transition (see Fig. 4).

The discrepancy between the model eq. 1 and the experimental data are much more evident in Fig. 7. The deviation from linearity $\Delta M$ was obtained after subtraction of the linear part $M = 5/6\chi_\perp H$ from the experimental data and from the fit data, respectively, (drawn as the black line in Fig. 5). Moreover, it is clear that the linear section in $M(H)$ data recorded for 577 K, which extends from 0 to 1.9 T, is a few times longer than the linear part of the fit covering range from 0 to 0.5 T. Therefore we deduce that the model equation (eq.1) is correct for the intermediate field range. On contrary, for the low fields there probably exists another mechanism or a different energy barrier for spin reorientations that stabilizes the spin cycloid. However, the exact nature of this field-induced transition is unknown. In case of the measurement carried out at 577 K, a deviation of the $\Delta M$ experimental data from linear dependence appears above the field $H_a$ which means that the spin-cycloid starts to be anharmonic (Fig. 6.b) leading to a nonlinear magnetization. Thus we deduce that magnetic field induces the following sequence of the transitions: in low field there is the circular spin-cycloid and BiFeO$_3$ compound exhibits the linear magnetization on the applied field; above the field $H_a$ the spin cycloid becomes anharmonic which causes the nonlinear magnetization and above the field $H_c$ almost all spins are aligned according with field direction and the system once again exhibits the linear magnetization $M(H)$.

The measurements of $M(H)$ magnetization curves and the anomalies taking place at $H_c$ and $H_a$ for various temperatures allowed us to propose the diagram presented in Fig. 8. Compared to the previous report by Tokunaga et al. [31] this diagram contains data from magnetization measurements close to $T_N$ and the additional line $H_a(T)$ associated probably with field-induced transformations of the cycloid. In the diagram, $H_c(T)$ and $H_a(T)$ lines separate different regimes in the spin arrangements related to the circular, the anharmonic and the homogeneous spin order.

However, the diagram proposed by us and the exact mechanism of the field-induced transition taking place at $H_a$ needs both experimental confirmation by means of various techniques (for example EPR, NMR and neutron diffraction) and also theoretical studies. So far, neutron diffraction experiments were performed at zero magnetic field for which the spin system exhibits the circular cycloid order in a whole temperature range. The same problem occurs in NMR measurements [36-38] performed in absence of an applied magnetic field because only the circular cycloid order is tested in this case. On the other hand, the EPR spectra recorded by Ruette et al. [32] in the wide magnetic field range (0-25 T) reveal a complex structure with few anomalies. Unfortunately, these authors have not analyzed the low-field anomaly, which could correspond to the possible transition from the circular to anharmonic spin order suggested by us.

**3.2 Crystal structure and thermal lattice expansion**

The analysis of X-ray diffraction of the sample of BFO ceramics preserved the rhombohedral (space group *R3c*) in accord to literature data [4]. This symmetry remains unchanged in the wide range of temperature, from 300 K up to 900 K. The X-ray pattern of the BFO obtained at room temperature is shown in Fig. 9. In addition to the main phase, traces of impurities are observed (noted by asterisks in Fig. 9). This impurity has been identified as Bi$_2$O$_3$. The content of this precipitation is about 4 %, estimated from the formula:

$$c = I_{prec}/(I_{prec} + I_{BFO}) \qquad (2)$$

where $I_{prec}$ and $I_{BFO}$ denote the most intensive lines belonging to the precipitation phase and to the main phase spectra, respectively. The relative volume expansion of the cell within the temperature range 300-900 K estimated, as $(\Delta V/V_0)\cdot 100\% \approx 2\%$ is quite small. One can see in Fig. 10 that increase of the lattice parameter *a* is insignificant since the relative expansion $(\Delta a/a_0)\cdot 100\% \approx 0.08\%$, while the lattice constant *c* changes markedly $(\Delta c/c_0)\cdot 100\% \approx 0.8\%$. It indicates that lattice parameter *c* is mainly responsible for expansion of the unit cell. There is no structural anomaly in vicinity of the magnetic AFM-PM phase transition occurring at $T_N$. However, one can distinguish a slight change in the rate of the crystal lattice expansion $(\Delta c/c_0)$. This structural transformation corresponds to the appearance of the magnetic phase transition in BFO (see Fig. 4).

### 3.3 Electrical properties

The effective dielectric permittivity $\varepsilon'(T,f)$ (Fig. 11.a) and the dielectric loss coefficient $tan\delta(T,f)$ (Fig. 11.b), show slight anomalies in vicinity of the temperature $T_N$ where the transition between the AFM and PM phases occur. The $\varepsilon'(T,f)$ increases markedly in the high temperature range, i.e. above the temperature $T_N$ where the magnetic transition exists. The dielectric loss coefficient value varies from about $tan\delta \approx 1$ at room temperature that remains in agreement with literature data [13] and reaches $tan\delta \sim 100$ in the 650-700 K range. Moreover, the $\varepsilon'(T)$ and $tan\delta(T)$ exhibit dispersion in the whole temperature range (300-700 K).

The ac electric conductivity $\sigma(T, f=100Hz)$ (Fig.12.a) temperature dependence is shown in the $T\sigma$ vs. $T^{-1}$ plot, according to the small polaron model proposed for the BiFeO$_3$ [18, 19]:

$$\sigma(T) = \sigma_0 T^{-1} exp(E_a/k_B T) \qquad (3)$$

The small polaron dependence (eq.3) is fulfilled below the temperature $T_N$ (see straight-line segment in Fig. 12.a). One can distinguish a change in the slope of the electric conductivity plot in vicinity of the magnetic phase transition $T_N$ temperature and the $T\sigma(T^{-1})$ curve deviates from the presumed dependence above the $T_N$ (Fig. 12.a). However it should be noted, that the thermally activated dependence, i.e. $\sigma \sim T^{-1}$ did not also fit the conductivity within this temperature range.

Hence, the derivative $d(\ln(T\sigma))/d(T^{-1})$ was calculated (Fig. 12.b). The horizontal part of the plot visible in the temperature range below ~640 K and above ~670 K confirms the small polaron mechanism of the electric conduction described with the formula (3). Therefore, the estimated activation energy equals to $E_a = 0.81$ eV below $T_N$. One can estimate $E_a \approx 0.54$ eV above 670 K, when assuming the small polaron model is satisfactorily approximated in the high temperature range.

The steep change in the $d(\ln(T\sigma))/d(T^{-1})$ value occurs within a several degrees range and the minimum in this plot anomaly occurs at $T_A = 650$ K. This anomaly in the electric conduction temperature dependence corresponds to the anomalies in the dielectric permittivity $\varepsilon'(T)$ and $\tan\delta(T)$, which occur within the 640-660 K range. Moreover, both the dielectric permittivity and the dielectric loss coefficient $\tan\delta(T,f)$ show dispersion. Such effects indicates an existence of the electric charge carriers, or space charge subsystem. Hence, the thermally generated charge carriers contribute to the measured values of the $C_p(T,f)$ and $G(T,f)$ quantities. Therefore we deduce that the anomaly in the electric properties, which manifest within the 640-660 K range, corresponds to the anomaly in the magnetic properties, which occur in the vicinity of $T_N = 646$ K (compare Figs 11 and 12 to Fig. 4).

On the other hand, the magnetic phase transition, manifested by the magnetization $M(T)$ peak detected near the Néel temperature occurring within several degrees i.e. $T_N \pm 8$ K range (Fig. 4), corresponds to a lack of structural, or crystal lattice parameters, and anomaly in this range. However, it worthwhile to note that a smooth change in the thermal expansion can be discerned (see Fig. 10). Hence, concerning the multiferroic features of the $BiFeO_3$ there is no direct interaction between the ferromagnetic and ferroelectric ordering that would lead to the multi-ferroic phase transition at the same critical temperature, since the values of $T_N = 646$ K and $T_C = 1143$ K are different.

We propose to ascribe the detected coincidence, between the magnetic and electric anomalies in vicinity of the Néel temperature $T_N$ in the $BiFeO_3$, to the contribution of the electric charge carrier subsystem. This effect can be described using the chemical potential approach [39, 40]. The possibility to register kinks in the temperature dependence of the conductivity of the investigated sample is a consequence of the chemical potential critical behaviour. The thermodynamic equilibrium condition requires that the chemical potentials of the sample $\mu_s$ and the chemical potential of the electrodes $\mu_e$, which are attached to enable the electrical measurement should be equal $\mu_s = \mu_e$. It has been shown [39, 40] that the chemical potential of a solid state material exhibits critical behaviour in the case of second order phase transitions. Therefore, the critical behaviour of the chemical potential $\mu_s$, related to the magnetic phase transition, should influence the flow of the electron gas in the electrode-sample-electrode system, when the electric conduction temperature dependence is measured. This is due to the fact that the chemical potential enters the theoretical Kubo formula [41] for the electrical conductivity. Consecutively, the change in the activation energy value reflects the transformation in the electronic structure when the phase transition between the ferromagnetic and the paramagnetic phases occurs in the $BiFeO_3$. Hence, the observed correspondence can be ascribed to the electronic subsystem and chemical potential features however a detailed discussion needs further studies.

## 4. CONCLUSIONS

To summarize, we report on the magnetic and dielectric measurements of the properties of the hot-pressed $BiFeO_3$ ceramic.

In magnetic properties, behavior of this sample is similar to that one exhibited by several single crystals [24] with negligible irreversible processes and the almost identical magnetization dependence on temperature $M(T)$ and hysteresis loops $M(H)$. Near the Néel temperature $T_N = 646$ K there appears a pronounced jump in a magnetic moment, which is found in single crystals also [24]. The reason for this feature is still unknown.

In the low magnetic field range there is an anomaly in $M(H)$, probably due to the field-induced transition from circular cycloid to the anharmonic cycloid. Some evidence for such transition in the form of low-field anomaly can be found also in the report by Ruette et al. [32] on EPR measurements but it was not analyzed and explained yet. At high field limit we observe the field-induced transition to the homogeneous spin order [25, 30, 33]. Therefore, we propose possible *H-T* diagram of the magnetic order in the bismuth ferrite, which contains the cycloidal, anharmonic and homogeneous magnetic orders. We suggest that magnetic field induces the following sequence of the transitions: (1) at low field there is a circular spin-cycloid and $BiFeO_3$ compound exhibits the linear magnetization on the applied field dependence; (2) above the field $H_a$ the spin cycloid becomes anharmonic which causes nonlinear magnetization, and (3) above the field $H_c$ almost all spins are aligned according with the field direction and the system once again exhibits linear magnetization $M(H)$. This proposal however, ought to be confirmed by other experimental methods and explained theoretically.

The anomalies in the electric properties, $\varepsilon'(T)$, $\tan\delta(T)$, and $\sigma(T)$, which are manifested within the 640-680 K range, coincide to the anomaly in the magnetization $M(T)$ dependence, which occurs in the vicinity of $T_N = 646$ K. We propose to ascribe this coincidence to the contribution of the electric charge carriers subsystem. This effect in the $BiFeO_3$ material can be described using the chemical potential formalism since the critical behaviour of the chemical potential $\mu_s$, related to the magnetic phase transition.

**Acknowledgements**

This project has been founded by National Science Centre (project No. N N507 229040). The discussion with dr. A. Rachocki from Institute of Molecular Physics PAN is kindly acknowledged.

**FIGURE CAPTIONS**

Fig. 1 ZFC and FC magnetization dependence on temperature $M(T)$ recorded for the applied magnetic field $\mu_0 H = 1$ T.

Fig. 2 Real $\chi'$ and imaginary $\chi''$ components of magnetic susceptibility measured for AC magnetic field at amplitude $H_{ac}=3$ Oe and frequency $f = 1000$ Hz.

Fig. 3 Antiferromagnetic hysteresis loops recorded for several temperatures: 4, 50, 100, 200, and 300 K.

Fig. 4 Magnetization peak showing a maximum which occurs near the Neel temperature $T_N = 646$ K ($T_N$ value has been determined at the half of the maximum amplitude).

Fig. 5 Magnetization dependence vs. magnetic field $M(H)$ for two chosen temperatures 304 and 577 K. $M_s$ is the spontaneous magnetization extrapolated from high-field range, denoted with a dashed line. $H_c$ is the critical field. The solid lines represent the best fits to the linear regimes occurring in the low field.

Fig. 6 The magnetic order in the spin cycloid. (a) the cycloid is: circular for low fields; (b) the cycloid becomes anharmonic in the intermediate fields above $H_a$ field; (c) the cycloid is destroyed above the critical field $H_c$, and the homogeneous order is established. $\boldsymbol{P}_s$ denotes here the vector of electric spontaneous polarization and $\boldsymbol{q}$ the propagation vector of the cycloid.

Fig. 7. Experimental data and the fit within the model equation (eq.1) after the subtraction of the linear part of the magnetization characteristic for the low field range. The $H_a$ denotes the transition field to the anharmonic order and the $H_c$ the critical field and transition to the homogeneous order.

Fig. 8 The proposed H-T diagram for the magnetic order within the spin cycloid.

Fig.9. The XRD pattern of the $BiFeO_3$ ceramics at 300 K.

Fig.10. Lattice parameters *a* and *c* of $BiFO_3$ vs. temperature.

Fig.11. (a) The dielectric permittivity temperature dependence $\varepsilon'(T)$ measured at $f$ = 0.10, 0.12, 1, 10, 20, and 100 kHz. (b) The dielectric loss coefficient $tan\delta(T,f)$ shown in vicinity of the magnetic phase transition temperature $T_N$ = 646 K.

Fig 12. (a) The electric conduction dependence $T\sigma$ vs. $T^{-1}$ plotted according to the assumed small polaron model. (b) The electric conductivity derivative $d(ln(T\sigma))/d(T^{-1})$ plot, the horizontal straight-line segments indicate the ranges proper for evaluation of the activation energy value.

**Figures**

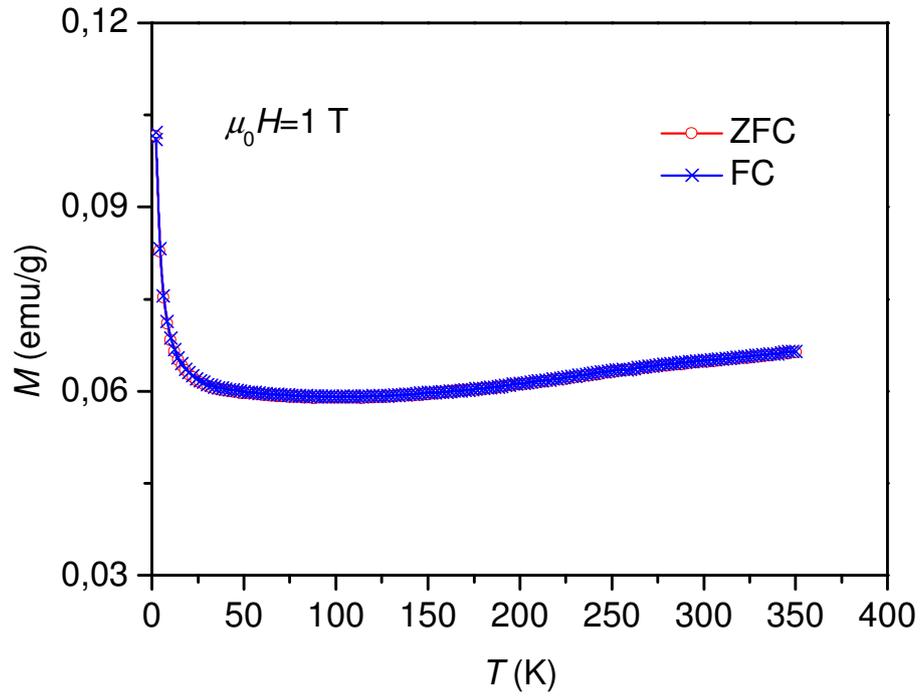

Fig. 1

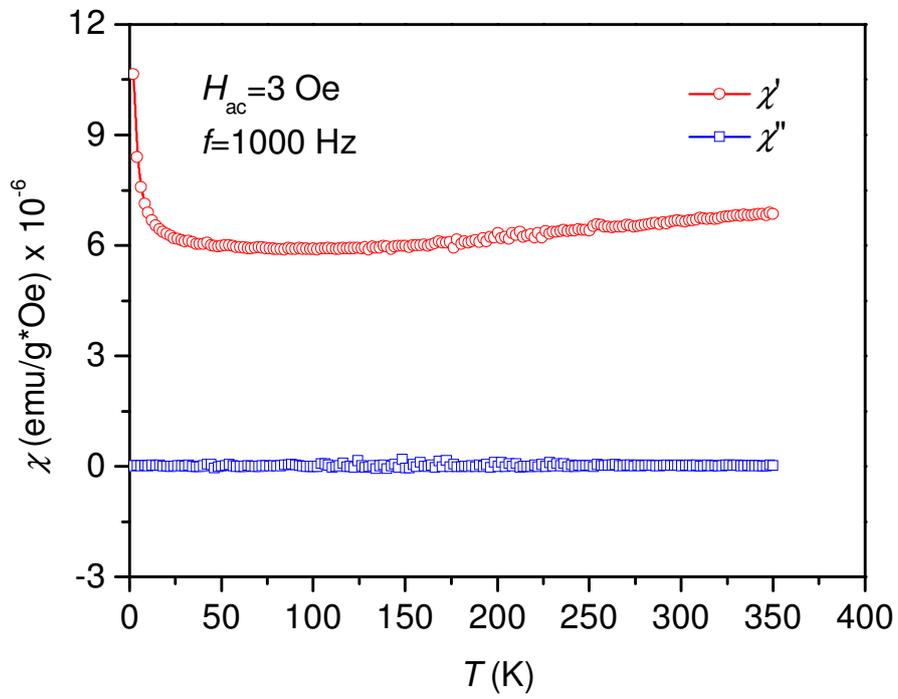

Fig. 2

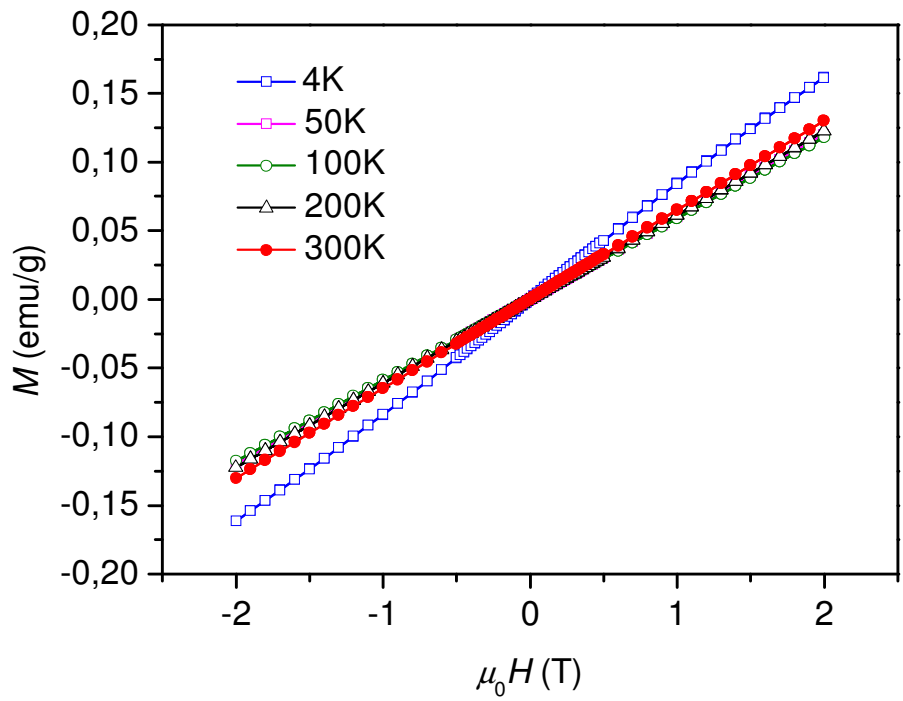

Fig. 3

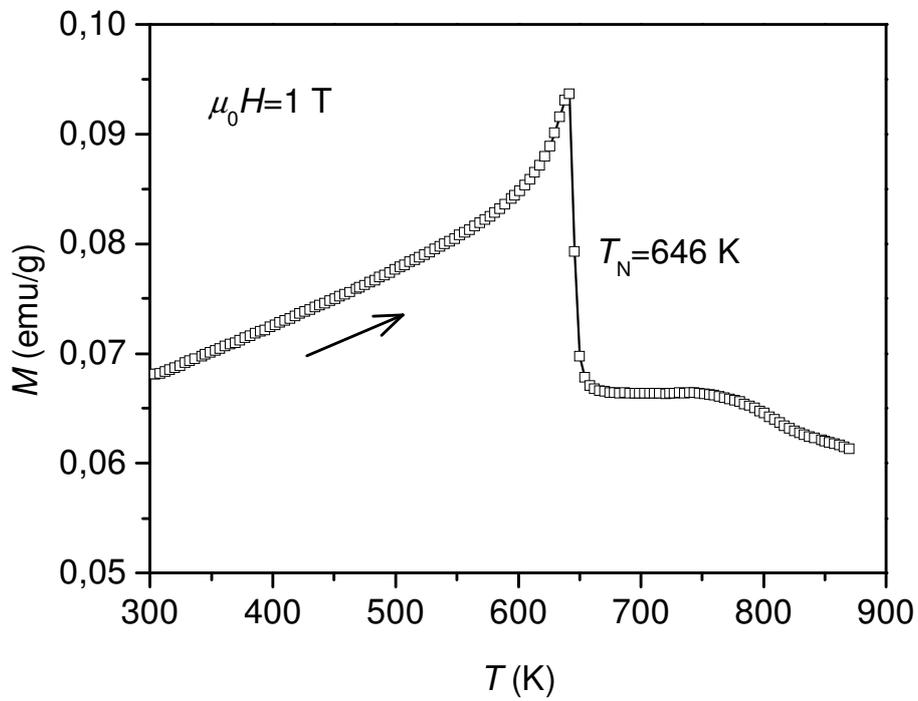

Fig. 4

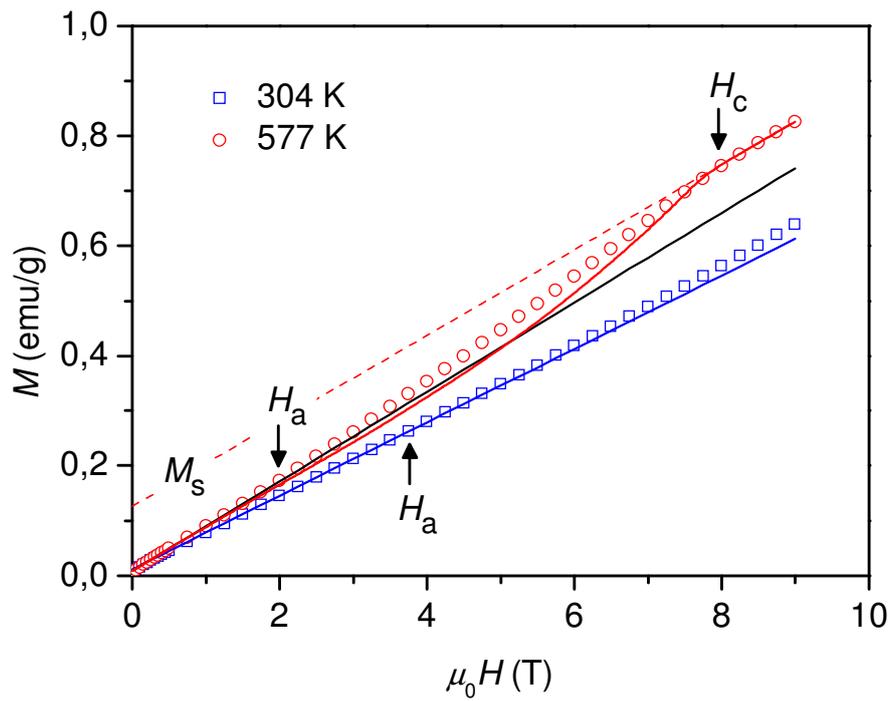

Fig. 5

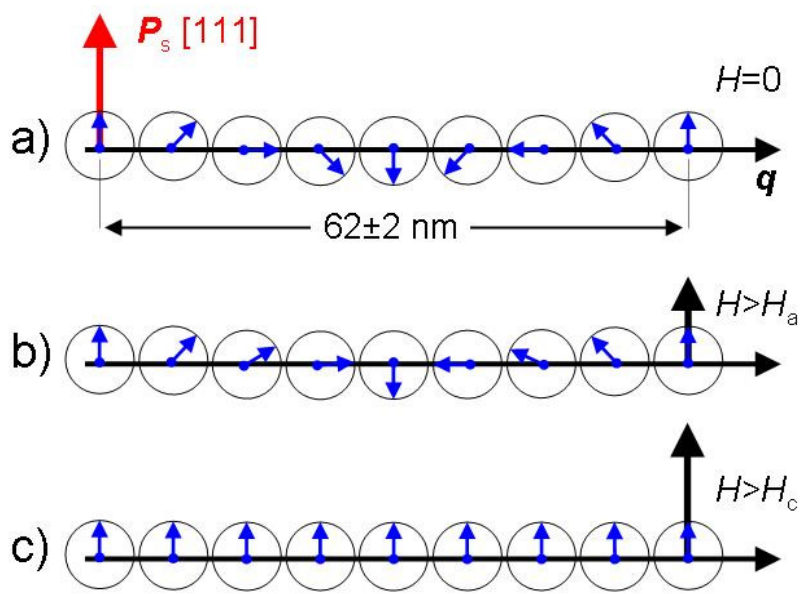

Fig. 6

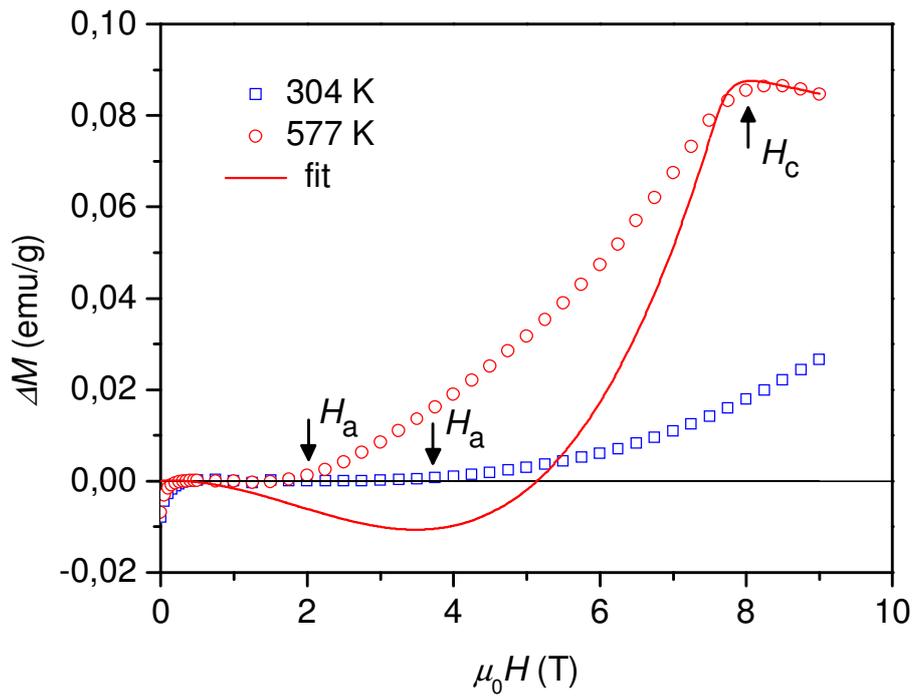

Fig. 7

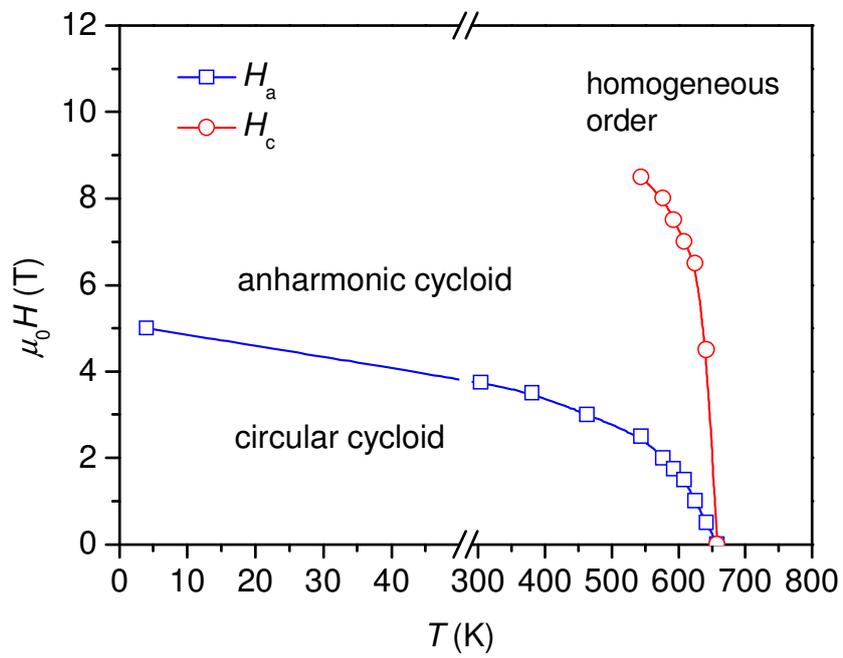

Fig. 8

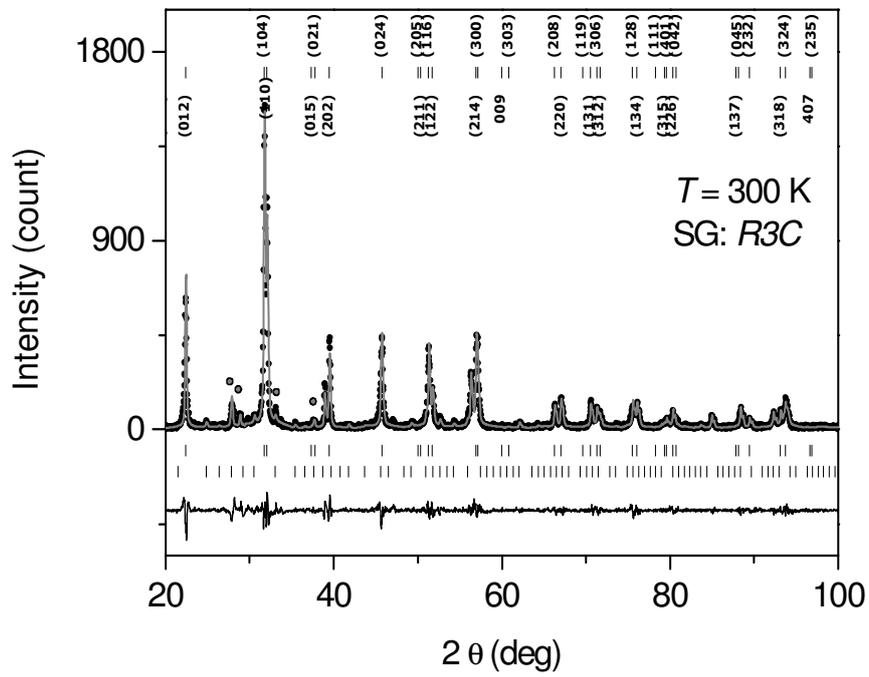

Fig 9. XRD pattern of the BiFeO$_3$ ceramics, obtained at 300 K

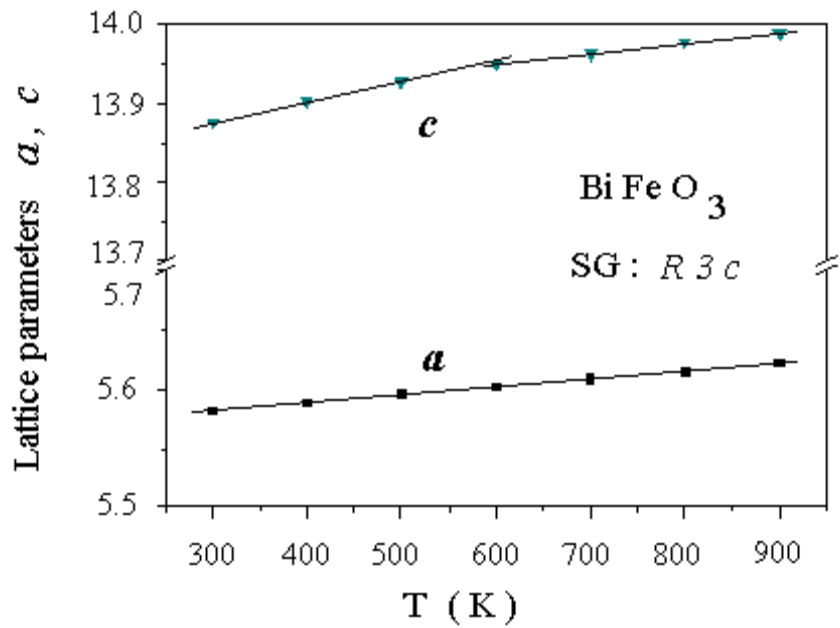

Fig.10 Lattice parameters of BiFO *vs.* temperature

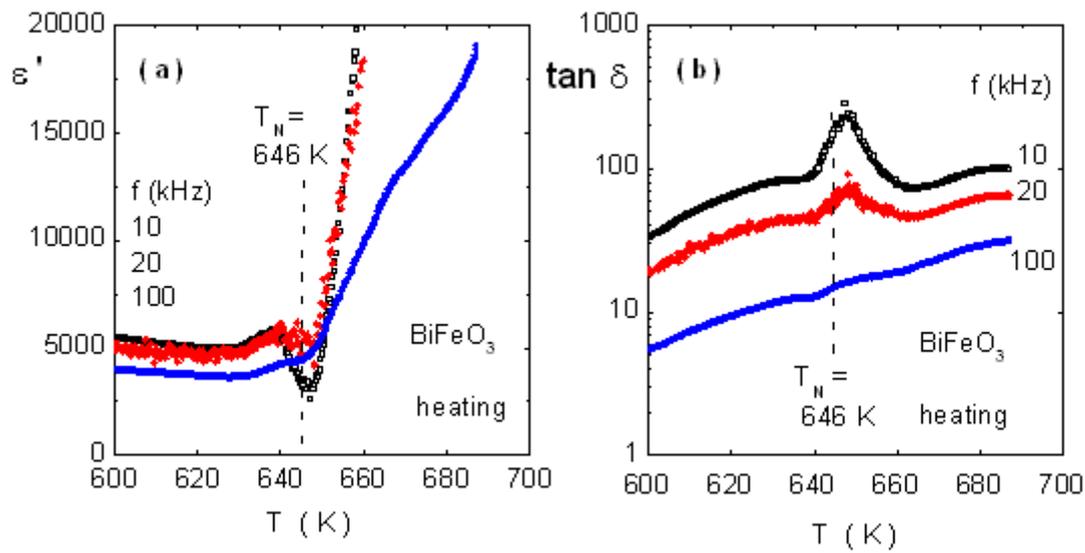

Fig. 11.(a) and (b)

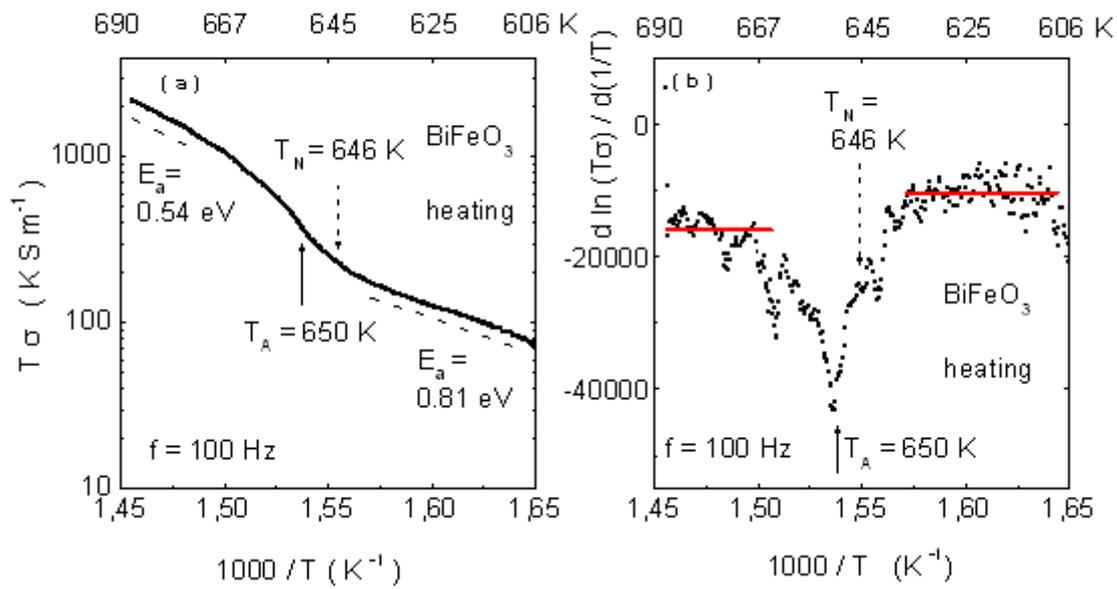

Fig. 12.(a) and (b)